# Technology Independent Method for defining Battery Powered LPWAN End-Point Lifetime Estimation

Adrià Galin-Pons
Universitat Autònoma de Barcelona
Edifici Enginyeria, Campus UAB
08193 Bellaterra, Spain
+34 65 988 1226
adria.galin@e-campus.uab.cat

## ABSTRACT
Low Power Wide Area Networks (LPWAN) are expected to play an important role in the deployment of the Internet of Things (IoT). One of the major advantages of LPWAN is the low power consumption characteristics which allow devices to be deployed without the restriction of power supply when battery is contained, in consequence the Lifetime estimation metric is of high relevance. This article provides a method for evaluating the Lifetime Estimation of battery powered LPWAN End Points, a depiction of the LPWAN architecture together with a breakdown of second order effects of the model are analyzed.

## 1. INTRODUCTION

The number of connected or "smart" devices that make up the Internet of Things (IoT) is rising day by day as more and more different connection technologies become available on the market. Several publications state the expected total number of connected devices is in the order of billions according [1] and [2], what seems to be clear is the expectations of massive deployment of IoT [3] in the next decade.

The IoT is based on a range of available technologies and/or families [4], i.e Sigfox, NB-IoT, LTE-CatM, etc. where each of them present different properties which allow different parameter optimization and features. The IoT interface connected via wireless links seems to be an attractive approach since the wireless communication provides less deployment costs and greater mobility characteristics; which will impact on the need of supplying power to the electronics by any other mean since continuous access to the power grid will be limited and not always guaranteed.

There are several methods to supply energy to an IoT device in order to allow the electronic system to operate as intended, one possible solution is to have the electronics itself connected to the power grid. Other approach is to have a battery powered system or rather having an energy harvesting scheme.

Since by definition, wirelessly connected devices present attractive deployment costs profiles driven by the non-restriction of a required connection to power supply, the battery approach is widely used. As a consequence, testing the power efficiency and the estimated battery lifetime of the End-Point electronic system has become critical.

Nowadays there is no consensus on how to define the lifetime estimation of IoT devices. Since the lifetime highly depends on the activity of the device there is no standard way to compare the lifetime values for different devices and communication technologies. Moreover the estimation method is not yet standardized.

The battery lifetime estimation of a given battery powered device is nowadays a key factor in determining which device suits the given Low Power Wide Area Network (LPWAN) application best. Due to the high demand, today there are several approaches that are used for determining the battery lifetime estimation with height variability, where in most of the cases the provided approximations are technology dependent which causes differences in the results and biased conclusions due to the calculation assumptions when comparing different End-Points. To achieve a higher level of homogeneity in the results, a common method for evaluating the battery lifetime estimation is required.

Nevertheless, the use of common methods for defining a way to determine the expected battery lifetime has not been fully addressed often overlooked in the literature, not considering the limitations that poor models could impose on the estimations. In the present paper it is going to be proposed an empirical approach to measure the power consumption for providing an accurate prediction of operational lifetime of a single-use battery LPWAN device.

The battery lifetime concept has been historically addressed by several standardization organizations of relevance for other wireless technologies and has therefore been discussed in most standardization committees worldwide during the last decade highlighting the inclusion of new requirements for ultra-low power devices during the releases number 12 and 13 of the 3GPP regarding cellular technologies, where it was first defined the power save mode (PSM) and enhanced DRX (eDRX) mode for reducing power consumption in those cellular devices. In LPWAN domain, still there is no common way to define consumption and/or battery lifetime considering the unique characteristics of this particular technology.

The present paper is organized as follows: we describe the LPWAN architecture in Section II and the elements constituting the network are introduced. Section III presents the lifetime estimation model depicting all relevant parameters for an accurate estimation. During the IV section the relevance of the self-discharge parameter are analyzed. In section V it is proposed different traffic models for evaluating the lifetime of a given device being technology independent. Section VI, before concluding, it is reviewed the second order effects not included into the model as well as its implications.

## 2. LPWAN ARCHITECTURE

While the impressive number of IoT-LPWAN technologies available on the market have many differences in their approaches and spectrum access -cellular, unlicensed, radio interface, etc-, the network topology structure is in common in all available families and technologies, in terms of start network topology. This sub-clause gives the reference architecture for an LPWAN system by means of a high-level decomposition into major components with a characterization of the interaction of the components itself.

The following figure depicts a LPWAN reference architecture in terms of functional blocks -i.e. the components- and their interfaces.

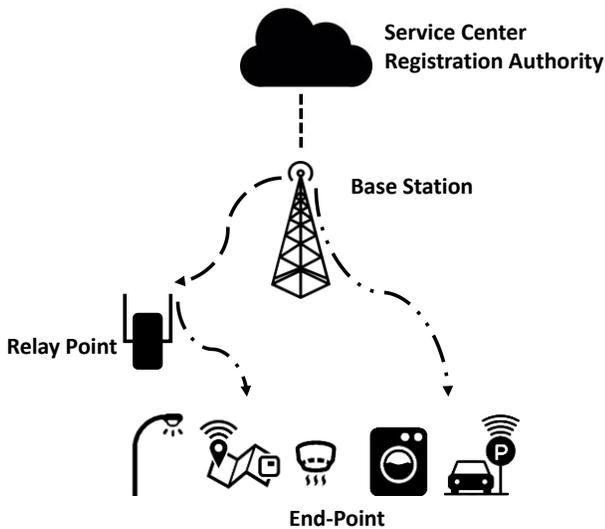

**Figure 1 LPWAN reference architecture**

The End Point is the element in the network that obtains the data and acts as a sensor and/or actuator in a LPWAN system. The end point connects the application data between the local device application and network application, and does have associated an identifier that is unique across the given LPWAN family, with this identifier the network is able to distinguish the given End Point with the other elements being present in the network. End Point aims to connect wirelessly with Base Stations and/or Relay Points from Fig1. with the aim of exchange the data. End Points can present mobility features being connected to different Base Stations through time, while those under a given Relay Point are limited to only communicate with that given Relay Point, unless otherwise reconfigured. The End Point may have specific Radio Interface with given spectrum access mechanism. The end Points may be battery powered devices and those are the subset under evaluation in the present paper.

The Relay Point is a static node that relays application data for a small number of end points (EPs) in a given geographical area. Relay Points may also communicate application data between its local device application and network application. Relay Points usually require external power, and, just like End Points, they are recognized by an identifier that is unique across all LPWAN families. They are particularly used to provide service to remote areas for a subset of End Points.

The base station are the radio hub of LPWAN systems. Base stations does have several interfaces, first of all they aim to connect with the service center (SC) as stated in Fig 1. This interface usually will not have wireless interface. Base Stations also connects with the end point (EP), and relay point (RP); those last interfaces are wireless communications and its channel access methods are LPWAN family dependent. Base stations may transmit significant amount of RF power in order to cover as much area as possible, in order to reduce the deployment costs.

Service Center and Registration Authority are the entities that aim to authenticate the End-Points within the Network, and they are internet based. They do not impact directly the End-Point Lifetime estimation.

## 3. LIFETIME ESTIMATION

Batteries consist of one or more electrochemical cells converting chemical energy directly to electrical energy. One of the major key features of a given Battery is its capacity, i.e. the number of electrons it can provide, which is defined by Eq, 1.

$$Q = I \cdot t \qquad (1)$$

Where $Q$ is the charge in coulombs, $I$ is the current in amperes and $t$ is the time in seconds. A single electron has 1.602e-19 coulombs of charge. Another important value is the energy stored in the battery E as defined by Eq. 2.

$$E = Q \cdot V \qquad (2)$$

Where $E$ is the energy stored in Joules, $Q$ is the charge in coulombs, and $V$ is the average voltage provided by the battery during discharge. For convenience, industry usually works with Ampere·hour units for charge, and Watt·hours for energy.

In order to achieve the requirements of the industry of expected battery lifetimes of the order of several years, IoT devices shall have tight requirements for power consumption in order to achieve the expected goals. There are several ways to reduce the consumption, reducing the duty cycle of the End-Point may provide a reduction of consumption as stated by the Authors in [4], in that publication it is stated that reducing the Duty Cycle will not only provide larger lifetimes but also will enable other services and network capacity improvements.

In order to verify the power consumption and expected battery lifetime, a new unified method for calculating the the LPWAN End-Point Batery Lifetime is presented in the present section.

The energy consumption of a device, is driven by several contributions: transmission, reception, computation, standby and self-discharge as defined in Eq. 3.

$$E_{total} = E_{tx} + E_{rx} + E_{proc} + E_{stby} + E_{loss} \quad (3)$$

Where:
$E_{tx}$ is the energy used for transmitting;
$E_{rx}$ is the energy used for receiving;
$E_{proc}$ is the energy used for processing;
$E_{stby}$ is the energy while being in standby mode;
$E_{loss}$ is the energy wasted by the intrinsic battery consumption.

Those contributions are the result of the typical execution phases of IoT End-Points. For the correct operation of the system they commonly are in standby mode most of the time, waiting for a trigger to activate as defined by the duty cycle of the system. The power consumption in standby mode is usually reduced significantly as processors go to deep low power modes that typically power off all the subsystems rather than the real time clock (RTC) and other basic circuits. Typical power consumptions in this state are under µWatts.

When the device is activated it usually performs some computation and potentially transmits (Tx) or receive (Rx) information from the network.

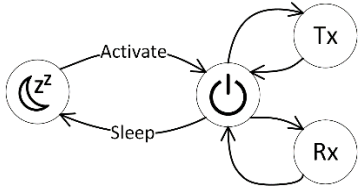

**Figure 2 State chart of a typical IoT operation**

IoT end-devices are typically using half-duplex communication, so they hardly ever transmit and receive simultaneously. Thus, power consumption can be modeled for each of the different phases depicted in Figure 2.

During communications the processor usually is not very active, however it is running in a higher power mode state than standby. The additional factor contributing to the energy consumption is the battery self-discharge.

Once the $E$ is defined and its contributions are identified from Eq. 3, the lifetime can be extracted by the relationship between energy and power as stated in Eq. 4.

$$t = \frac{E}{P} \quad (4)$$

## 4. POWER MODELS

Let us suggest in other terms the equations as stated above, where the continuous operation of the device under evaluation shall be divided in operational modes grouped by its consumption magnitudes in order to obtain accurate consumption measurements essential for accurate lifetime estimations.

### 4.1 Duty cycle

We would suggest an activation cycle of the device (a period in seconds) being $t_{activation}$ which must be analyzed prior of measuring. Once defined, there is a probability of computing, involved transmitting and receiving activities, being able to measure the power of each activity independently and with a significant improvement on the accuracy.

$$\propto_{comp} + \propto_{tx} + \propto_{rx} + \propto_{idle} = 1 \quad (5)$$

$$\propto_{idle} = 1 - (\propto_{comp} + \propto_{tx} + \propto_{rx}) \quad (6)$$

### 4.2 Communications

In this part we are going to measure the power associated to the communication states as defined above: transmission and reception. The Transmission shall include the related payload associated to the transmission, signaling and acknowledgements if applicable. Those parameters will be technology dependent. Power here includes the communication module and the processor power consumption during transmission, which is running in a power mode higher than standby.

Once the value $P_{tx}$ is obtained, there is a clear relationship between $E_{tx}$ as stated in the next equation:

$$E_{tx} = \frac{P_{tx} \propto_{tx} t}{t_{activation}} \quad (7)$$

The transmission energy could also be defined by a message size in bits (including headers). $S_{msg}$ to be sent in every activation.

Then the Bitrate $B_{tx}$ can be included into the Eq. 7 as defined in Eq. 8.

$$E_{tx} = \frac{P_{tx} S_{tx\,msg} t}{B_{tx} t_{activation}} \quad (8)$$

Where $P_{tx}$ is the power consumption of transmitting 1 bit. However former equation can be enough for a simplified analysis.

The same procedure shall be applicable for reception mode. In analogy as above, once the $P_{rx}$ is measured, there is a relationship with the related energy $E_{rx}$ as stated in the following equation:

$$E_{rx} = \frac{P_{rx} \propto_{rx} t}{t_{activation}} \quad (9)$$

### 4.3 Computation

It is also essential for an accurate lifetime estimation to consider the computational energy costs. The computational costs may be present when accessing data from sensors, or doing some mathematical computation. This is the time when the processor is active but not transmitting.

The Power consumption here will be mainly driven by the dynamic power consumption of CMOS circuits.

$$E_{proc} = \frac{P_{dyn} \propto_{proc} t}{t_{activation}} \quad (10)$$

where $P_{dyn}$ can be depicted as:

$$P_{dyn} = \propto_{dyn} C_{eff} f V^2 \quad (11)$$

where $\propto_{dyn}$ is the activity factor of the transistors of the circuit, $C_{eff}$ is the total effective capacity of the circuit, $f$ is the frequency of operation of the processor, and $V$ is the power supply Voltage of the processor.

In order to reduce the computational energy costs, there are only two available options: rather reduce the operating voltage, which will be limited by the operational voltage threshold or rather reduce the battery capacity. This last topic is limited by the number of operation to be processed.

### 4.4 Idle

In order to enhance the battery lifetime the idle -or standby- mode is used for reducing the consumption of the End-Point. This mode allows the device to become in a state of utra-low power current drain which may result in an optimization of the leakage power associated in this mode of operation.

The measurement procedure for determining the idle consumption consist on measuring $P_{stby}$ and then extract the related energy $E_{stby}$ as shown in the next equation:

$$E_{stby} = \frac{P_{stby} \propto_{stby} t}{t_{activation}} \quad (12)$$

### 4.5 Self-Discharge

The self-discharge effect refers to the Intrinsic Battery Consumption, this part is mainly Battery and Application - temperature, humidity and atmospheric pressure- dependent. Due to the Chemical response of the Battery itself it loses its stored Energy capacity through time. This point must be considered and has a relevant impact when dealing with large lifetime periods as the case of LPWAN. It is well known that batteries by the electrical-chemical responses its stored energy E as defined in Eq. 2 is time dependent, and due to external effects the stored energy decreases with time.

Battery self-discharge rate is typically reported by industry as a percentage of energy loss per months. This value is known as $D$. In the following equation it is stated the $E_{loss}$ equation as a function of $D$.

$$E_{loss} = E_{total}(1 - D)^{t_{months}} \quad (13)$$

### 4.6 Complete Model

Considering the previous contributions, the total Energy model can be further developed from Equation 3 with the aim of obtaining the relationship within the time dimension.

$$E_{total} = E_{tx} + E_{rx} + E_{proc} + E_{stby} + E_{loss} \quad (14)$$

When replacing the $E_{loss}$ by its definition the following equation can be deduced to estimate the battery lifetime in years of a generic LPWAN End Point as:

$$t(1-D)^{\frac{t}{k_{spm}}} = \frac{t_{activation}}{(P_{tx}\propto_{tx} + P_{rx}\propto_{rx} + P_{proc}\propto_c + P_{idle}\propto_{idle})} \quad (15)$$

where $k_{spm}$ is the number of seconds in a month.

At this stage time variable can be extracted so lifetime can be finally determined by the empirical parameters we are able to measure. The Lifetime equation is defined as following:

$$t = \frac{k_{spm} \, W\left(\frac{t_{activation}\log(1-D)}{(P_{tx}\propto_{tx} + P_{rx}\propto_{rx} + P_c\propto_c + P_{idle}\propto_{idle})k_{spm}}\right)}{\log(1-D)}$$

$$(16)$$

where $W$ is the Lambert $W$ function.

Other researchers also end up considering the Lambert W function when estimating the battery model as indicated in [5]. Lambert W function is the inverse of $f(x) = xe^x$.

## 5. EMPIRIC MEASUREMENT

Lifetime depends basically of two main blocks as identified in the present paper. The Device consumption and the Intrinsic Battery Consumption. In order to perform the associated testing according the previous clause, the measurement setup is presented below: for the aim of the measurement an oscilloscope is required with sufficient sample rate, the probe which will provide the power supply to the Device Under Test (DUT) as well as the acquisition of the consumption metric. Finally a controlling laptop is required which will control the measurement equipment together as the DUT for tuning it in an appropriate configuration and traffic model prior to test.

In the following figure it is illustrated the general measurement setup where power and measuring connections are depicted by continuous lines between the measuring equipment while intermittent interconnection means control buses typically by means of LAN, GPIB or USB interfaces.

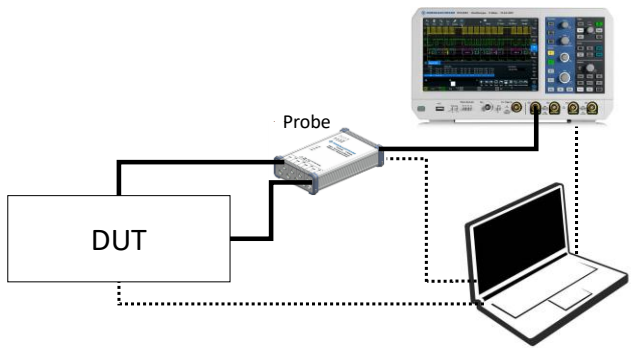

**Figure 3 Measurement setup for performing the empirical measurement**

The device consumption concept is basically related on how it operates and how it has been designed. There are several hardware items a developer shall take care when designing their electronic devices if long operating times are a requirement in order to reduce its associated consumption. The key parts of the circuits and subsystems includes: the microcontroller, the wireless controller, the sensor arrays, the power supplies inputs and outputs and any other specific subsystems that could potentially draw significant current; all those items are critical and shall be considered when dealing with multi-year expected lifelong devices.

The wireless technology in place used by the device (End Point) to communicate is clearly dependent on the consumption, the protocol being used, and the firmware, are also essential topics which may affect and modify the lifetime estimations. Therefore, when considering the consumption of the EPs four operational working states of the device are defined:

- Transmission (Tx) including payload, EP signalling and/or ACK if aplicable;
- Reception (Rx) if aplicable;
- Computation;
- Stand-by.

During each of the above states, the consumption shall be measured by means of a calibrated test equipment consisting of a voltimeter and amperementer with sufficient resolution and sample rate for providing accurate measurements with low uncertainty as shown in Figure 3. The measurement equipment shall make sure that even the very short power consumption peaks are captured.

It shall be noted that between the different states the predicted consumption may be significatively different, for this reason it is recommendable to check separately each of the above states. The amperemeter is typically equipped with internal shunt that shall be selected for one of the different ranges. A maximum resolution shall be needed to have amperemeter of the order of µA resolution. There are several available solutions on the marked including the R&S RT-ZVC Multi-Channel Power Probe.

## 6. SELF-DISCHARGE PARAMETER EFFECTS

When considering long life battery measurement, it is needed to consider the self-discharge of the battery itself, due to the intrinsic physics of the accumulator it loses capacity through time. Mostly the Battery Manufacturers refers to this value by percentage of capacity lost per month. According the state of the art Battery Manufacturers can provide batteries of a self-discharge parameter of 0.5%.

Assuming a state of the art $D$ of 0.5% self-discharge per month, during the first 4 years the battery has lost 20% of its stored Energy capacity without any transmission nor computation, just by self-discharge effect.

At a given time interval, according Eq. 13 the stored energy $E$ is calculated with different $D$ values, represented in Fig 4 it is noted the relevance of $D$ and how little deviation in this value can provide significant changes on the resulting lifetime estimation.

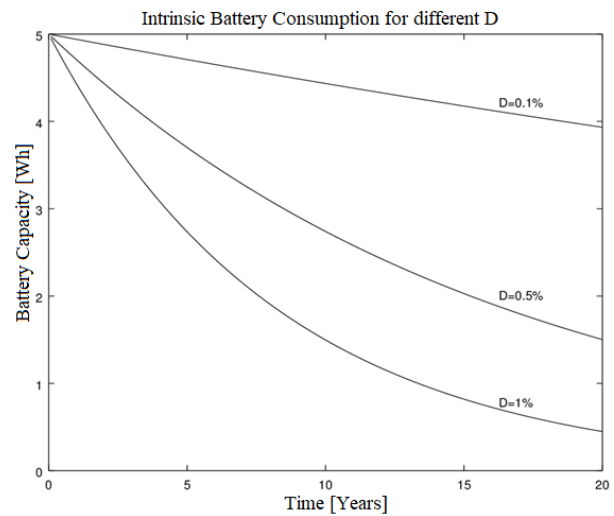

**Figure 4 Effect of $D$ parameter on Self-Discharge Curve for a generic 5Wh battery. Different $D$ has been considered**

In the Fig 4 it is represented the loss of the capacity in terms of energy E of the battery for different $D$ values. From the figure above can be observed that at the end of the eighth year of the battery it has lost 40% of its capacity assuming state of the art $D$=0.5%.

## 6.1 Linear vs Exponential Discharge Approach

Some articles and LPWAN players have provided an analysis considering linear discharge curve. A linear simplification seems that can be valid for a relative short lifetime. However as the target duration increases the difference between those two approaches becomes significant. It has to be noted that linear self-discharge curve is more strict in terms of intrinsic consumption than the exponential one, providing shorter lifetime estimations.

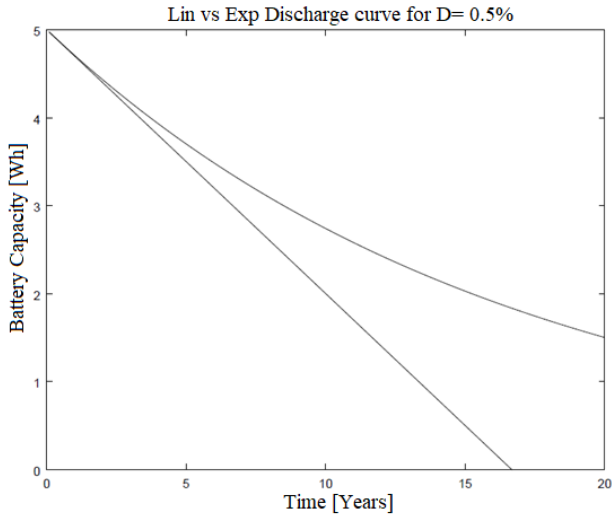

**Figure 5 Linear vs Exponential approach of Self Discharge model for 5Wh Battery with $D$ = 0.5%**

From the figure above it is evident that assuming a linear model with a state of art $D$ of 0.5% a given End Point will never have greater lifetime than 16 years. However, it is also noted that the linear discharge curve does not describe de discharge process of the battery accurately. An exponential self-discharge response, on the other hand, describes more precisely the behavior of the battery whatever is its related chemical composition – i.e. lithium ion, nickel cadmium, acid lead batteries, etc - providing accurate estimations. There is a clear need of considering exponential discharge curve models into the lifetime estimation for LPWAN battery powered devices.

## 7. TRAFFIC SCENARIOS

In order to compare the performance in terms of Battery Lifetime of two different LPWAN, it is essential to normalize at the same traffic model for comparison purposes. In the present section, three different traffic models have been defined covering relevant LPWAN use cases.

The eq. 16 does describe the lifetime estimation referred to values that can be empirically measured. For this very reason three End-Points traffic profiles has been identified: one transmission per day, one transmission per hour and ten transmission per hour. It is noted that each LPWAN family may have different approaches, including signaling, acknowledgements, etc. It is essential to define a common traffic model in order to compare lifetime estimation values between different LPWAN End-Points.

In the following table are depicted the set of parameters needed to measure according Eq.16 in order to obtain the battery powered End-Point lifetime estimation.

**Table 1. Parameters to measure**

| Parameter | Description | Units | Method |
|---|---|---|---|
| $P_{tx}$ | Transmission Power | Watts | Average Power during transmission |
| $P_{rx}$ | Reception Power | Watts | Average Power during reception |
| $P_{proc}$ | Computing Power | Watts | Average Power during computation |
| $P_{idle}$ | Idle Power | Watts | Average Power during idle |
| D | Self-Discharge Rate | % per month | Reported by manufacturer |

For a given Traffic Model (TM) the relationship between End-Point modes must be obtained by means of measuring at least one full cycle at a given Traffic Model resulting the measuring time the one defined in TM period. We present three different traffic models to be considered for the calculation of lifetime. As stated previously within the TM period interval it shall be performed the Transmission, Reception, computation and Idle modes separately in order to increase the accuracy of the measurement.

In Table 2 it is depicted the acquisition measurement times for the proposed traffic models.

**Table 2. Measurement time per traffic model**

| Traffic Model | Measurement Cycle |
|---|---|
| 1 Msg per Day | 1 Day |
| 1 Msg per Hour | 1 hour |
| 10 Msg per Hour | 6 min |

Being able to measure the different consumptions as defined above in Table1 including Transmission, Reception, computation and Stand-by modes by means of proper equipment as identified in Figure 3 in a given traffic mode (TM) will provide with lifetime estimation values.

## 8. SECOND ORDER EFFECTS

In this section second order effects are analyzed and depicted, those are not included into the model and may lead to biased estimations which may impact on the analysis of LPWAN end point lifetime estimation. There are some relevant second order effects which will be relevant when determining the battery lifetime of an LPWAN End Points as the chip voltage threshold and the non-ideal discharge response, both causing a reduction of the effective lifetime as stated in Eq. 16.

The first issue is fluctuations of circuit performance [7]. The performance fluctuation distribution, which is caused by fabrication process deviation, battery capacity degradation providing supply voltage variation, and temperature variation; this effect becomes large as consumption takes place and the battery supply voltage is reduced by the loss of Charge effect.

Another issue with second order effects is the sub-threshold leakage current. Low-power circuit designs usually scales down the size of the MOS devices to improve CMOS large-scale integration (LSI) achieving a supply voltage reduction for low power operations; however, in order to maintain the LSI performance under a low supply voltage condition may potentially cause high amounts of leakage current. Moreover the threshold voltage of the internal circuitry of the RF device is different according particular specifications, this difference reduce the operational margin and performance of the internal circuitry when operating under a low supply voltage conditions as stated by the authors in [8].

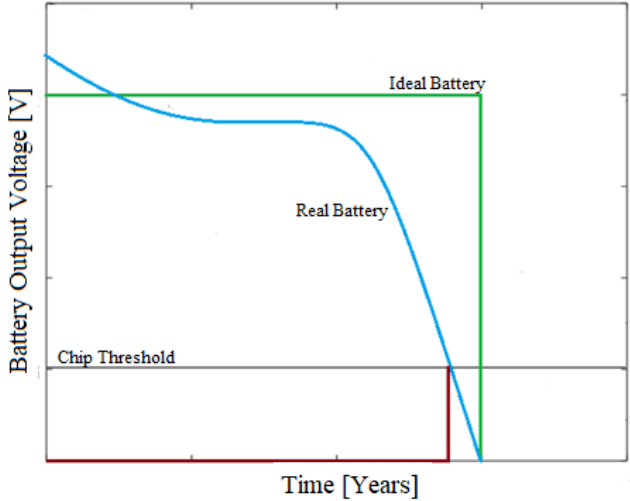

**Figure 6 Generic Real-Batery vs Ideal-Batery discharge curve**

When dealing with batteries together with an electronic system it is evident that the electronics in order to perform given operations they must be supplied at a nominal voltage. As the supply is being reduced at a given time the electronics stops its intended behavior, known as Chip threshold. Since the battery response which is dependent on the battery type –blue curve- cannot be modeled as ideal –green curve- lifetime may be reduced due to the mentioned second order effects.

The Medium Access Control (MAC) layer in Height Density Areas also plays an important role in the lifetime estimations. This point is not device dependent, but surrounding dependent. The MAC layer controls how different LPWAN devices share the given spectrum and ensures reliable packet transmissions [9]. In a high density environment the probability to find collisions between different LPWAN devices becomes significant and the need of retransmission can be required those forcing the LPWAN devices to retransmit given packets and increasing the required consumption accordingly.

Notwithstanding the proposed model as indicated in Eq. 16 may potentially be impacted by second order effects on the End Point lifetime estimation causing a reduction on the expected values.

## 9. CONCLUSIONS

This paper has reviewed the need of proposing a unique technology independent method for estimating the operation lifetime of single use battery-powered LPWAN. Moreover, this paper has reviewed different techniques and issues for enhancing the battery lifetime as well as we have presented a novel common method for determining the Battery Lifetime of an LPWAN End Point independently on the technology being used, and the type of battery in place. Second order effects and system level analysis have been introduced and summarized, but not considered into the model, it has been shown that the second order effects may impact on a reduction of the lifetime. Moreover, challenges related to common traffic models have been discussed with the aim to compare effectively the lifetime of different LPWAN families normalized at the same traffic model. The importance of self-discharge battery parameter has been compared, discussed and analyzed. Future research should consider the potential second order effects, moreover future test trials are necessary to validate the kinds of conclusions into the lifetime model that can be drawn from this publication.